\documentclass[aps,preprint,preprintnumbers,amsmath,amssymb,amsfonts,groupedaddress,superscriptaddress,showpacs,showkeys,floatfix]{revtex4-1}

\usepackage{amsmath}
\usepackage{amssymb}
\usepackage{bm}
\usepackage{graphicx}
\usepackage{xcolor}
\usepackage{epstopdf}

\newcommand{\bmath}{\begin{mathletters}}
\newcommand{\emath}{\end{mathletters}}
\newcommand{\be}{\begin{eqnarray}}
\newcommand{\ee}{\end{eqnarray}}
\newcommand{\ba}{\begin{array}}
\newcommand{\ea}{\end{array}}

\newcommand{\no}{\nonumber}

\begin{document}

\title{Electronic effect of doped oxygen atoms in Bi2201 superconductors determined by scanning tunneling microscopy}
\author{Ying Fei}
\author{Kunliang Bu}
\author{Wenhao Zhang}
\author{Yuan Zheng}
\affiliation{Department of Physics, Zhejiang University, Hangzhou 310027, China}
\author{Xuan Sun}
\author{Ying Ding}
\author{Xingjiang Zhou}
\affiliation{Beijing National Laboratory for Condensed Matter Physics, Institute of Physics, Academy of
Sciences, Beijing 100190, China}
\affiliation{Collaborative Innovation Center of Quantum Matter, Beijing 100871, China}
\author{Yi Yin}
\email{yiyin@zju.edu.cn}
\affiliation{Department of Physics, Zhejiang University, Hangzhou 310027, China}
\affiliation{Collaborative Innovation Center of Advanced Microstructures, Nanjing 210093, China}

\clearpage

\begin{abstract}

The oxygen dopants are essential in tuning electronic properties of Bi$_2$Sr$_2$Ca$_{n-1}$Cu$_n$O$_{2n+4+\delta}$ superconductors.
Here we apply the technique of scanning tunneling microscopy and spectroscopy to study the influence of oxygen dopants in an
optimally doped Bi$_2$Sr$_{2-x}$La$_x$CuO$_{6+\delta}$ and an overdoped Bi$_{2-y}$Pb$_y$Sr$_2$CuO$_{6+\delta}$.
In both samples, we find that interstitial oxygen atoms on the SrO layers dominate over the other two forms of oxygen dopants, oxygen vacancies on the SrO layers and interstitial
oxygen atoms on the BiO layers.
The hole doping is estimated from the oxygen concentration, as compared to the result extracted from the measured Fermi surface.
The precise spatial location is employed to obtain a negative correlation between the oxygen dopants and the inhomogeneous pseudogap.
\end{abstract}

\maketitle

\section{Introduction}
\label{sec1}

The copper oxide high-temperature superconductor is synthesized by doping a parent Mott insulator~\cite{LeeRMP06,keimerNat15}.
For Bi$_2$Sr$_2$Ca$_{n-1}$Cu$_n$O$_{2n+4+\delta}$ (BSCCO) superconductors, the extra oxygen
stoichiometry ($\delta$) is empirically adjusted by a controlled annealing in oxygen atmosphere. The disordered interstitial
O atoms introduce hole carriers to the CuO$_2$ plane. The holes thereafter play a primary role
in tuning electronic orders such as the superconducting and pseudogap (PG) states. In the phase diagram,
the superconducting state exists within a dome shaped regime, while the critical temperature
of the PG state is monotonically decreased with hole doping~\cite{MatsudaPRB99,HufnerRPP08}. At a fixed low temperature, the monotonic decrease
of the PG magnitude is observed as the hole doping is increased by the increase of O dopants~\cite{HufnerRPP08}.

The oxygen stoichiometry is in general difficult to  be predetermined quantitatively in annealed BSCCO samples.
Scanning tunneling microscope (STM) is a microscopic tool of probing samples at subatomic resolution~\cite{FischerRMP07}.
The in-situ measurement of differential conductance spectra is used to
explore the superconducting and PG states at a given spatial location~\cite{FischerRMP07}. A few abnormal spectral features
have been observed and attributed to the occupation of an O dopant in different forms~\cite{McElroySci05,ZeljkovicSci12,ZeljkovicNanoLett14}.
After an initial discovery of the interstitial O atoms on the BiO layers in Bi$_2$Sr$_2$CaCu$_2$O$_{8+\delta}$ (Bi2212) samples~\cite{McElroySci05},
two other forms of O dopants, the interstitial O atoms and O vacancies, were identified on the SrO layers~\cite{ZeljkovicSci12,ZeljkovicNanoLett14}.
The spatial distribution of O dopants of each form can be subsequently determined by a characteristic
differential conductance map. On one hand, the quantification of the density of O dopants allows an
estimation of the hole carrier density to be compared with results from other methods. On the other hand,
the precise location of O dopants helps a numerical analysis to extract a `local' relation between the
O dopants and the inhomogeneous PG state.
In Bi2212, the different relations were observed as the sample is changed from underdoping to optimal doping~\cite{ZeljkovicSci12}.

In addition to Bi2212, Bi$_2$Sr$_2$CuO$_{6+\delta}$ (Bi2201) samples belong to the other type of layered BSCCO materials.
Due to their crystal structural difference, the electronic properties of the two types of BSCCO samples are quantitatively different.
For example, the optimal superconducting critical temperature of Bi2201 is $T_c\approx32$ K, which is much lower
than the value of $T_c\approx90$ K in Bi2212~\cite{FengPRL02,EisakiPRB04}. The PG state in Bi2201 extends to the overdoped regime~\cite{HeSci14} while
the termination point of the PG state in Bi2212 is not conclusive yet~\cite{HufnerRPP08,KinodaPRB03,FujitaSci14}. Following the same
technical procedure of probing the O dopants, the studies of Bi2201 can thus verify both the reliability of the experimental
methodology and the applicability of the previous Bi2212 results to the whole BSCCO family.

In this paper, we apply STM to study the O dopants in an optimally doped and an overdoped Bi2201
sample. The spatially dependent differential conductance spectra are collected, accompanying the
topographic measurement over a clean area. The voltage is applied in a broad range to capture the
abnormal spectral features of O dopants. The conductance maps at specified voltages are extracted to
characterize the spatial distributions of the three types of O dopants.
The influence of the O dopants in a specific form to the electronic properties is investigated
through the calculation of the hole carrier density and the correlation with the PG distribution.
For both samples, we confirm a negative
correlation between the interstitial O atoms on the SrO layers and the inhomogeneous PG.

\section{Experimental Method}
\label{sec2}

From the material respect, the partial substitution of Sr by La, or Bi by Pb, has been applied to
produce pure single crystals of Bi2201 over a broad range of doping. %~\cite{MengSUST09,ZhaoCPL10}.
We select a Bi$_2$Sr$_{2-x}$La$_x$CuO$_{6+\delta}$ (La-Bi2201 with $x=0.37$ and $T_c\approx32$ K)
and a Bi$_{2-y}$Pb$_y$Sr$_2$CuO$_{6+\delta}$ (Pb-Bi2201 with $y=0.25$ and $T_c\approx13$ K)
to represent the optimal-doped and overdoped Bi2201, respectively.
The high-quality single crystals of La-Bi2201 and Pb-Bi2201
are grown by the traveling-solvent floating-zone method~\cite{MengSUST09,ZhaoCPL10}. Typical samples are cut
from the as-grown ingots and annealed with specific temperature and atmosphere conditions~\cite{MengSUST09,ZhaoCPL10}.

In our ultrahigh vacuum (UHV) STM system~\cite{zhengSciRep17}, single crystal samples are in-situ cleaved and inserted
to the STM head for a low temperature measurement at $T=4.5$ K. The STM topography is taken at a particular set
of the sample bias $V_\mathrm{b}$ and setpoint current $I_\mathrm{s}$. The local differential conductance ($dI/dV$) spectra, as
a function of voltage $V$, are acquired by a standard lock-in technique with
a modulation frequency of $f=983.4$ Hz. The scanning tips are etched electrochemically from tungsten wires,
and treated by electron-beam sputtering and field emission cleaning on a Au (111) crystal sample.
 For both La-Bi2201 and Pb-Bi2201 samples, a set of
representative data are shown as below.

\section{Results}
\label{sec3}

\begin{figure}[tp]
\centering
\includegraphics[width=0.55\columnwidth]{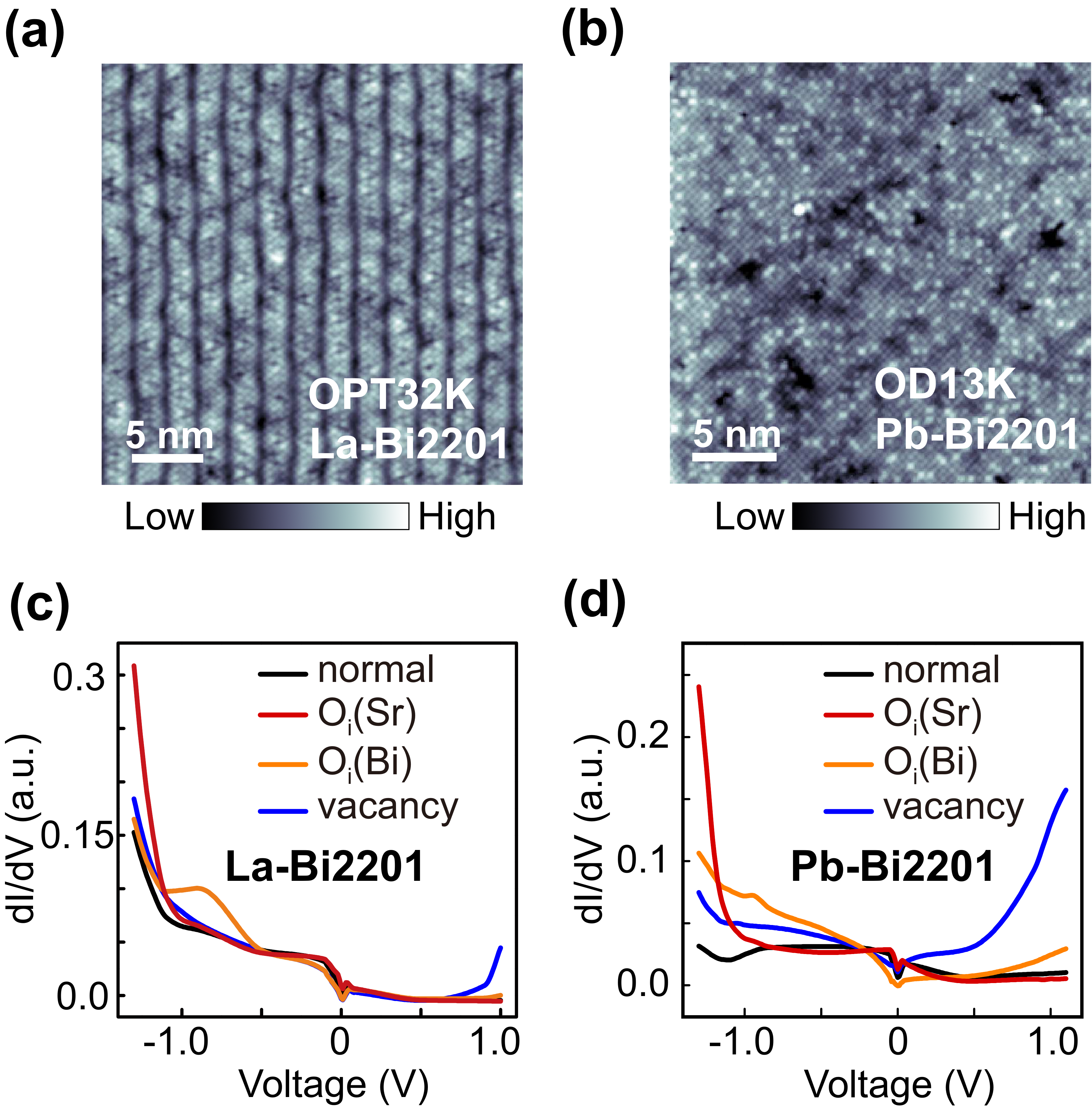}
\caption{(a-b) Two topographic images for (a) an optimally doped La-Bi2201 ($30$ nm $\times$ $30$ nm) and (d) an overdoped Pb-Bi2201 ($25$ nm$\times$ $25$ nm).
Both are acquired at the bias voltage of $V_\mathrm{b}=100$ mV and the setpoint current of $I_\mathrm{s}=100$ pA.
(c-d) The representative $dI/dV$ spectra for (c) La-Bi2201 and (d) Pb-Bi2201: two black lines are acquired at normal positions while
the lines in other colors are acquired at positions occupied by O dopants of three different forms.}
\label{fig:fign1}
\end{figure}

Figures~\ref{fig:fign1}(a) and~\ref{fig:fign1}(b) display two topographies taken on the exposed BiO layers
of La-Bi2201 (30 nm $\times$ 30 nm) and Pb-Bi2201 (25 nm $\times$ 25 nm), both containing a square lattice of Bi atoms. An additional supermodulation
structure is observed in La-Bi2201, while such a supermodulatation is completely suppressed in Pb-Bi2201
due to the elimination of a periodic potential of strain~\cite{MaoPRB93,SlezakPNAS08}.
In the same field of view (FOV) as in the two topographies, we take measurements of the differential conductance ($dI/dV$) spectrum
over a broad voltage range of $-1.3~\mathrm{V}\le V \le  1.0~\mathrm{V}$.
Figures~\ref{fig:fign1}(c) and~\ref{fig:fign1}(d) present four typical $dI/dV$ spectra in La-Bi2201 and Pb-Bi2201, respectively.
For the results of La-Bi2201 in Fig.~\ref{fig:fign1}(c), the red line exhibits
a sudden increase of $dI/dV$ with the decreased voltage for $V<-1.1$ V;
the blue line exhibits a sudden increase of $dI/dV$ with the increased voltage for $V>0.8$ V;
the orange line contains a resonance peak of $dI/dV$ appears around $V\approx-0.9$ V.
These three characteristic spectra correspond to atomic defects of different forms. As a comparison,
the spectrum at a normal position on the BiO layer is shown in the black line.
The $dI/dV$ spectra with similar features are found in Pb-Bi2201 (see Fig.~\ref{fig:fign1}(d)).
In previous studies of Bi2212 samples,
the three defect features were attributed to three different forms of O dopants~\cite{ZeljkovicSci12,ZeljkovicNanoLett14}.
Here we make the same assignment: the fast increase of $dI/dV$ below $V<-1.1$ V
for an interstitial O atom on the SrO layer, the resonance peak around $V\approx-0.9$ V
for an interstitial O atom on the exposed BiO layer, and the fast increase above $V>0.8$ V
for a missing O atom at its lattice site of the SrO layer. For conciseness, the first two forms
of defects are referred to as O$_{\rm i}$(Sr) and O$_{\rm i}$(Bi),
while the last form is named the O vacancy.

\begin{figure}[tp]
\centering
\includegraphics[width=0.75\columnwidth]{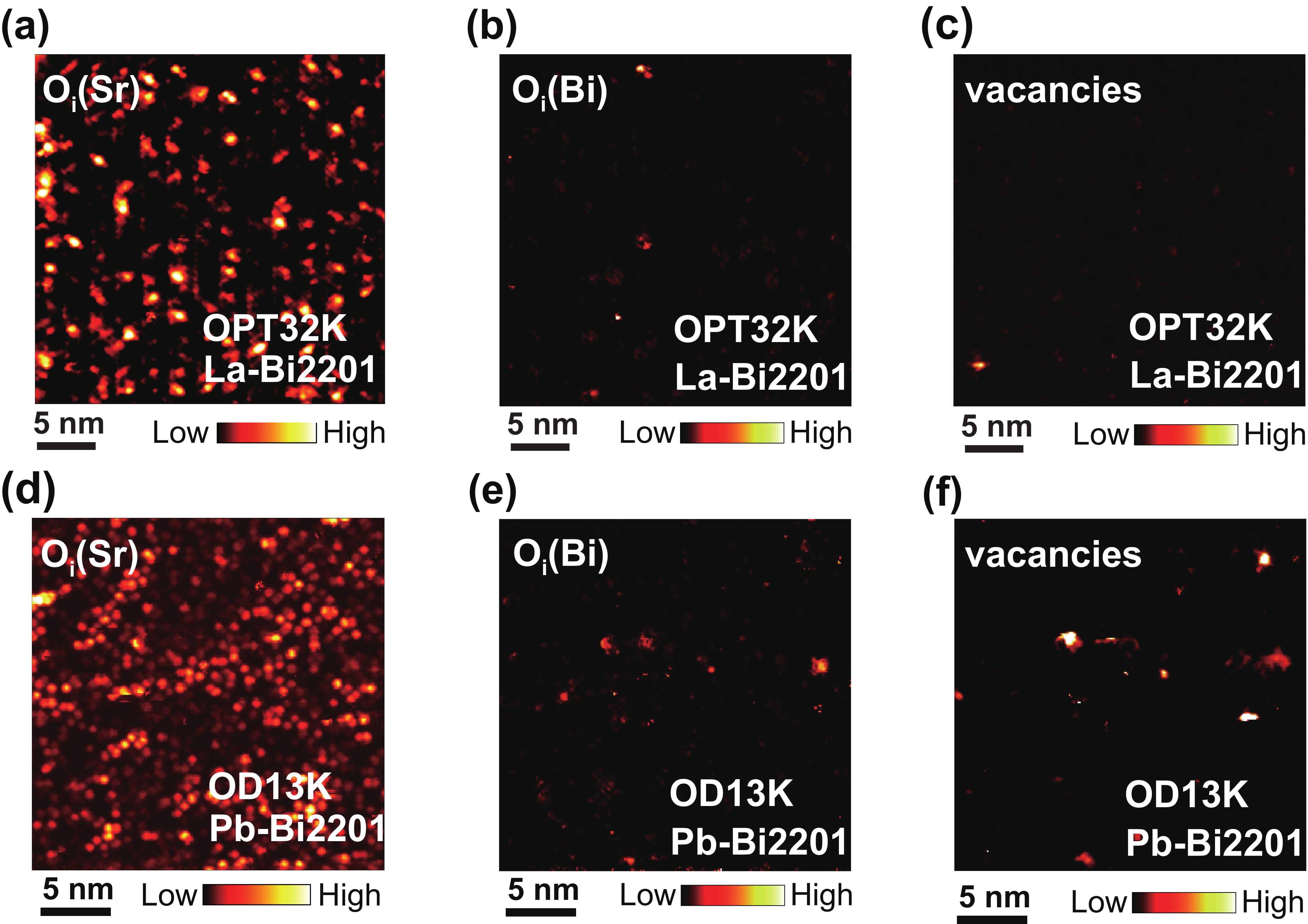}
\caption{The differential conductance maps characterizing O dopants. For La-Bi2201 (FOV in Fig.~\ref{fig:fign1}(a)),
the voltages are (a) $V= -1.3$ V, (b) $V=-0.9$ V and (c) $V= 1$ V.
For Pb-Bi2201 (FOV in Fig.~\ref{fig:fign1}(b)), the voltages are (d) $V= -1.3$ V, (e) $V=-0.9$ V and (f) $V= 1$ V.
}
\label{fig:fign2}
\end{figure}

The differential conductance maps of La-Bi2201 and Pb-Bi2201 at the voltage of $V=-1.3$ V
are displayed in Figs.~\ref{fig:fign2}(a) and~\ref{fig:fign2}(d), respectively.
In both conductance maps, a large number of atomic-scale bright spots appear on a dark background,
and each bright spot represents an O$_{\rm i}$(Sr) defect. As a comparison,
O$_{\rm i}$(Bi) defects and O vacancies are rarely observed
in their characteristic conductance maps (see Figs.~\ref{fig:fign2}(b)-\ref{fig:fign2}(c) and Figs.~\ref{fig:fign2}(e)-\ref{fig:fign2}(f)).
In our two Bi2201 samples, the O$_{\rm i}$(Sr) defects dominate over the other two forms of O dopants,
which is different from the case in Bi2212~\cite{ZeljkovicSci12}.
The O vacancies are gradually filled in Bi2212 with the increased doping,
and the density of O vacancies approaches to zero with the change
from the underdoped to optimally doped sample.  The two samples studied in this paper are
optimally doped La-Bi2201 and overdoped Pb-Bi2201 so that a negligible density of O vacancies
is expected. On the other hand, the density of O$_{\rm i}$(Bi) defects is a quite large number
in underdoped and optimally doped Bi2212.
The negligible density of O$_{\rm i}$(Bi) in both of our samples is possibly materially dependent~\cite{MaoPRB97}.

\begin{figure}[tp]
\centering
\includegraphics[width=0.55\columnwidth]{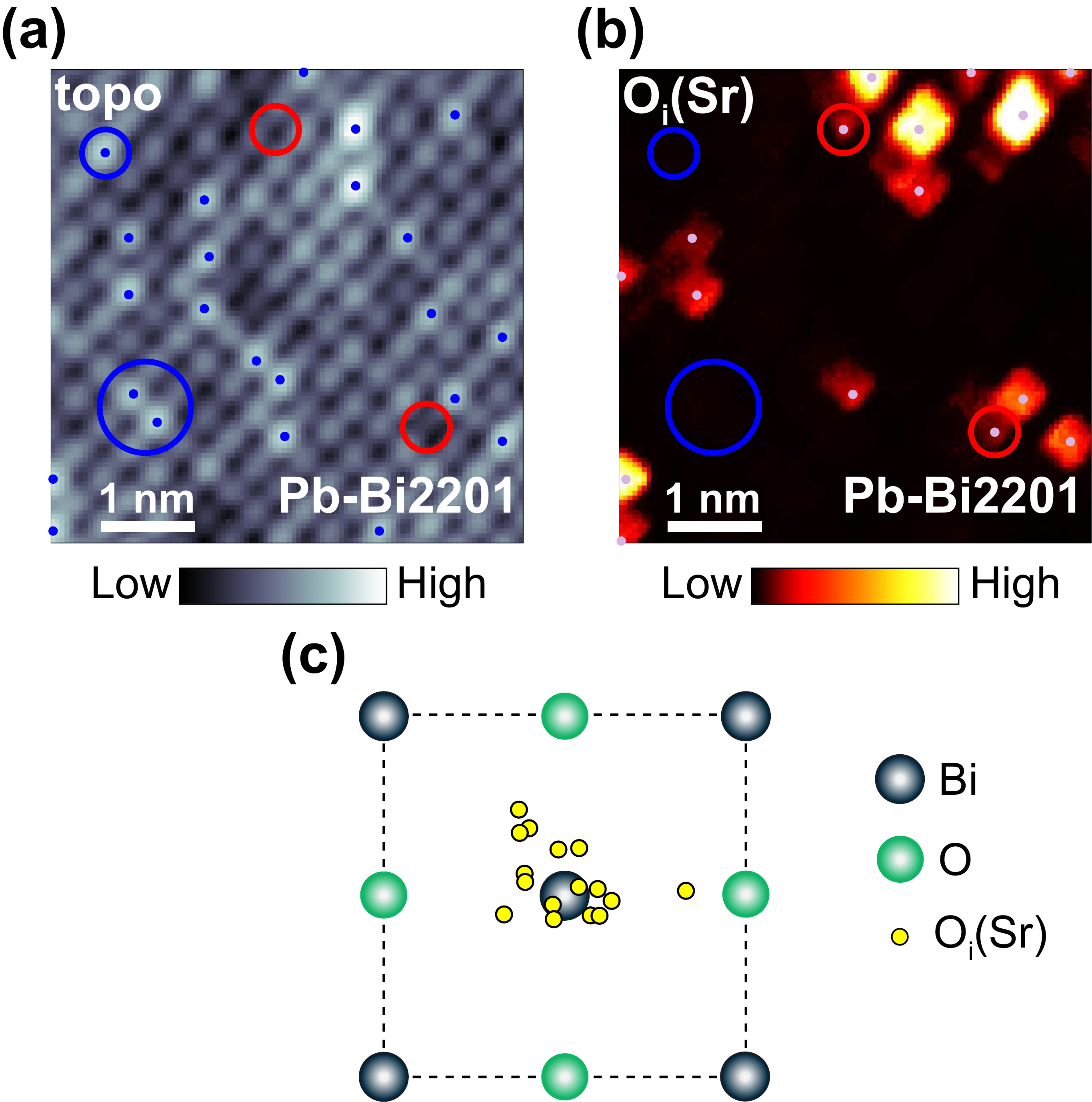}
\caption{
(a) A topography of Pb-Bi2201 (5 nm $\times$ 5 nm) acquired at $V_\mathrm{b}=100$ meV and $I_\mathrm{s}=50$ pA.
The Pb atoms substituting for the Bi atoms are marked by blue dots.
(b) The $dI/dV$ map at $V= -1.3$ V acquired in the same FOV as in (a). The O$_{\rm i}$(Sr)
defects are marked by white dots. The Pb atoms and O$_{\rm i}$(Sr)
defects do not appear simultaneously within the blue and red circles in (a) and (b).
(c) The scatter plot of the relative spatial locations of the O$_{\rm i}$(Sr) defects
in the BiO lattice.
}
\label{fig:fign3}
\end{figure}

To further clarify the spatial locations of O$_{\rm i}$(Sr) defects in Pb-Bi2201, we measure
the topography and differential conductance map ($V=-1.3$ V) simultaneously
in a smaller 5 nm $\times$ 5 nm area than that in Figs.~\ref{fig:fign1}(b) and~\ref{fig:fign2}(d)
but with a higher spatial resolution of $\sim0.5$ \r{A}, as shown in
Figs.~\ref{fig:fign3}(a) and~\ref{fig:fign3}(b). The centers of bright spots (marked by white dots) in the conductance map  are
precisely located on the measured topography of the BiO layer. With respect to the primitive cell of
the Bi-O lattice, the relative positions of all the O$_{\rm i}$(Sr) defects are recorded in a scatter plot as shown in Fig.~\ref{fig:fign3}(c).
Most of these defects are confined to the Bi sites instead of the O sites. Since the STM measurement only
determines the $(x, y)$ coordinates, the O$_{\rm i}$(Sr) defects are expected to be located vertically below
the Bi sites on the lower SrO layer. On the other hand, the substituting Pb atoms in general occupy
the Bi sites of the exposed BiO layer and can be identified
by isolated bight spots (marked by blue dots) in the topography.
The comparison between the Pb atoms in the topography and the O$_{\rm i}$(Sr) defects in
the conductance map shows a mild correlation. These two defects can be spatially independent from each other.
For example, only one feature (either
Pb atoms or O$_{\rm i}$(Sr) defects) appears within the two blue and two red circles in
Figs.~\ref{fig:fign3}(a) and~\ref{fig:fign3}(b). Our experiment thus suggests that the characteristic
spectral feature of $V<-1.1$ V is not Pb induced~\cite{KinodaPRB05}.

\begin{figure}[tp]
\centering
\includegraphics[width=0.55\columnwidth]{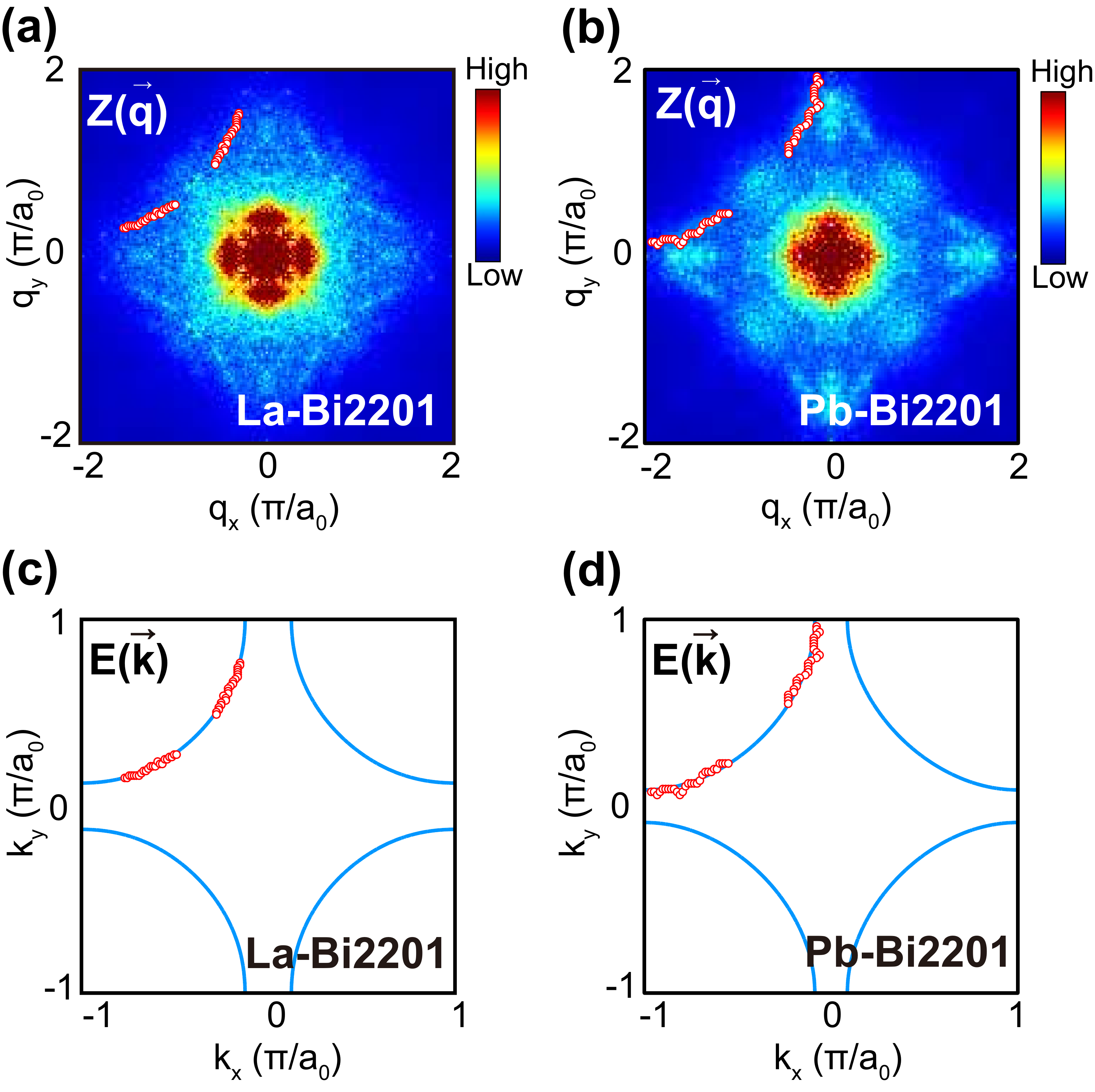}
\caption{(a-b) The integrated Fourier transformed ratio $Z(\boldsymbol q)$-maps for (a) La-Bi2201 (FOV in Fig.~\ref{fig:fign1}(a))
and (b) Pb-Bi2201 (FOV in Fig.~\ref{fig:fign1}(b)).  Open red circles mark the scattering momentums $\boldsymbol q$
of strong QPI signals. In the first Brillouin zones of the $\boldsymbol k$-space for (c) La-Bi2201 and (d)  Pb-Bi2201,
the Fermi momentums $\boldsymbol k_{\rm F}$ extracted from the QPI $\boldsymbol q$-momentums are also shown in open red circles.
The solid lines in (c) and (d) are the numerical fitting results of the entire FS following a tight-binding model in Eq.~(\ref{eq02}).
}
\label{fig:fign4}
\end{figure}

In the family of BSCCO materials, the O dopants directly introduce hole carriers whose concentration
is a fundamental parameter. As the number of interstitial O atoms is quantified in
the conductance maps, the hole doping can be estimated by twice the density of O dopants. For our La-Bi2201 and Pb-Bi2201
samples, this crude estimation leads to $p \approx 0.11$ and $p \approx 0.26$, respectively. However, we notice that the estimated
hole carrier concentration in optimally doped La-Bi2201 is smaller than the value of $p=0.16$ used in the empirical Presland~\cite{PreslandPhysC91}
and Ando~\cite{AndoPRB00} formulas. To understand this difference, we apply an alternative method of estimating the hole doping
with the measurement of the FS~\cite{HeSci14,FujitaSci14}. In the Luttinger theorem~\cite{LuttingerPR60}, the hole doping level $p$
is given by
\be
p=2\frac{A_\mathrm{FS}}{A_\mathrm{BZ}}-1,
\label{eq01}
\ee
where $A_\mathrm{FS}$ is the area of the hole pocket centered at $(\pi, \pi)$
and $A_\mathrm{BZ}$ is the area of the square-shaped first Brillouin zone (BZ).
In the STM experiment, the FS can be constructed from quasiparticle
interference (QPI) patterns in the $dI/dV$ maps of $g(\boldsymbol{r},V)$~\cite{FujitaSci14,kohsakaNat08,McElroyNat03,HanaguriNatPhys07,HeSci14}.
In practice, the ratio map of $Z(\boldsymbol{r},V)=g(\boldsymbol{r},V)/g(\boldsymbol{r},-V)$
and its Fourier transformed map of $Z(\boldsymbol q, V)=\mathrm{FT}[Z(\boldsymbol r, V)]$ are applied,
which enhances the QPI signal and cancels a setpoint effect~\cite{HanaguriNatPhys07}.
In the $Z(\boldsymbol q, V)$-map, the maximum QPI intensities
arise from elastic scattering between high density of states (DOS) regions in the momentum $\boldsymbol k$-space.
Figures~\ref{fig:fign4}(a) and~\ref{fig:fign4}(b) display the two integrated ratio maps of
$Z(\boldsymbol q)=\sum_V Z(\boldsymbol{q},V)$ for optimally doped La-Bi2201 (FOV in Fig.~\ref{fig:fign1}(a))
and overdoped Pb-Bi2201 (FOV in Fig.~\ref{fig:fign1}(b)), respectively.
The integration is over the voltage range of 5 mV $<V<$ 20 mV so that QPI patterns
around the Fermi level can be efficiently collected~\cite{HeSci14,FujitaSci14}. As shown in Figs.~\ref{fig:fign4}(a) and~\ref{fig:fign4}(b), each trace of
the scattering $\boldsymbol q$-wavevectors forms a closed path centered at $(\pi,\pi)$.
Following the octet model, the scattering $\boldsymbol q$-wavevector is twice the normal-state Fermi momentum $\boldsymbol k_{\rm F}$,
i.e., $\boldsymbol k_{\rm F}=\boldsymbol q/2$~\cite{HeSci14,FujitaSci14}. In the momentum $\boldsymbol k$-space, the scattered data of
$\boldsymbol k_{\rm F}$ acquired from the QPI measurement are plotted in Figs.~\ref{fig:fign4}(c) and~\ref{fig:fign4}(d).
To draw the entire FS, we introduce a tight-binding model~\cite{HeRHSci11}, where energy dispersion curves are expressed as
\be
  \varepsilon({\boldsymbol k})&=&\varepsilon_0+2t_0\left[\cos({k_x}a_0)+\cos({k_y}a_0)\right]\no \\
  & & +4t_1\cos({k_x}a_0)\cos({k_y}a_0)+ 2t_2\left[\cos(2k_xa_0)+\cos(2k_ya_0)\right] \no \\
  & & +4t_3\left[\cos({2k_x}a_0)\cos(k_ya_0)+\cos(k_xa_0)\cos(2k_ya_0)\right] .
  \label{eq02}
\ee
Following the approach in Ref.~\cite{HeSci14}, we fix four parameters, $t_0=-0.22$ eV,
$t_1=0.034315$ eV, $t_2=-0.035977$ eV and $t_3=0.0071637$  eV, and allow the reference
energy $\varepsilon_0$ to be varied. The FS is extracted self-consistently by fitting
the solution of $\varepsilon(\boldsymbol k_{\rm F})=0$ with the QPI-determined Fermi
momentums. Figures~\ref{fig:fign4}(c) and~\ref{fig:fign4}(d)
present the best fitting results of $\boldsymbol k_{\rm F}$ with $\varepsilon_0=0.211$ eV for La-Bi2201
and $\varepsilon_0=0.246$ eV for Pb-Bi2201. With the area
$A_{\rm FS}$ of the hole pocket calculated subsequently, the hole carrier concentration
is estimated by Eq.~(\ref{eq01}), giving $p\approx 0.21$ for La-Bi2201 and $p\approx 0.28$ for Pb-Bi2201.
For overdoped Pb-Bi2201, the values of hole doping estimated from the density of O dopants and the
Luttinger count are close to each other. For optimally doped La-Bi2201, the two estimations are however
different, one below the number of 0.16 in the Presland and Ando formulas and the other above.
The substitution of Sr by La (or Bi by Pb) may affect the density of admitted O dopants for as-grown samples and
partially contribute some carriers~\cite{EisakiPRB04,FujitaPRL05}.
In fact, the same disparity between the two estimation methods
can be found from the data of Bi2212~\cite{ref}. The quantification of the hole doping is thus
a nontrivial problem, which requires theoretical improvement to correct the oversimplified assumptions in
these two methods~\cite{BerlijnPRL12}.

\begin{figure}[tp]
\centering
\includegraphics[width=0.55\columnwidth]{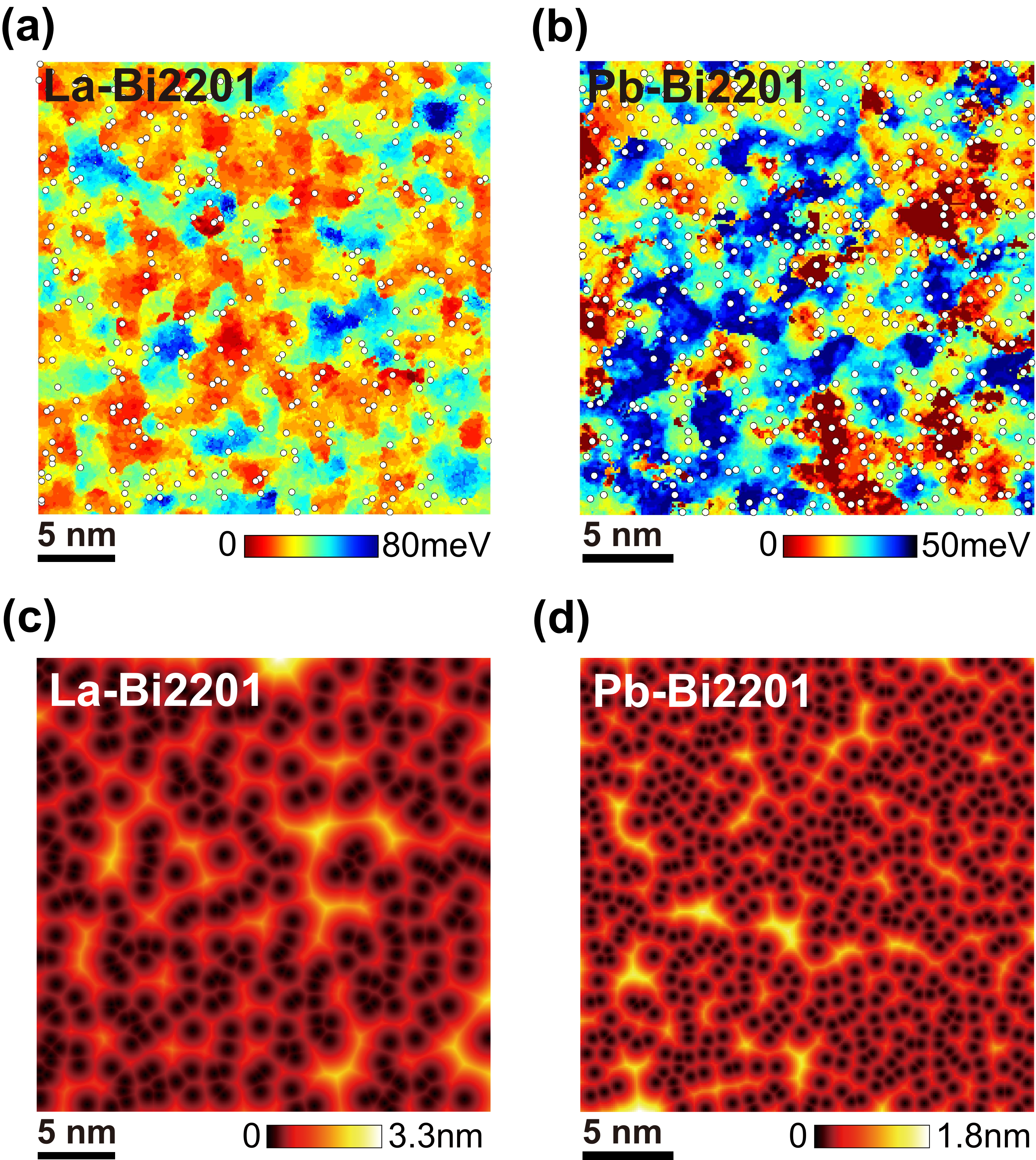}
\caption{(a-b) The maps of the PG magnitude for (a) La-Bi2201 (FOV in Fig.~\ref{fig:fign1}(a)) and (b) Pb-Bi2201 (FOV in Fig.~\ref{fig:fign1}(d)).
The locations of the O$_\text{i}$(Sr) defects determined from Figs.~\ref{fig:fign2}(a) and~\ref{fig:fign2}(d) are marked by white dots.
(c-d) The maps of the distance away from the nearest O$_\text{i}$(Sr) defect for  (c) La-Bi2201  and (d) Pb-Bi2201.}
\label{fig:fign5}
\end{figure}

The STM technique with a subatomic resolution allows a delicate investigation of the microscopic
relation between the O dopants and the electronic orders such as the PG states.
At each spatial location, the STM measures an individual $dI/dV$ spectrum, from which a `local' PG magnitude of $\Delta(\boldsymbol r)$ can be extracted.
The spatial distribution of the PG magnitudes is then depicted in a two-dimensional  map.
In Figs.~\ref{fig:fign5}(a) and~\ref{fig:fign5}(b), we show the PG maps of La-Bi2201 (FOV in Fig.~\ref{fig:fign1}(a))
and Pb-Bi2201 (FOV in Fig.~\ref{fig:fign1}(b)), respectively. A strong inhomogeneity appears in both PG maps~\cite{PanNat01,LangNat02,BoyerNatPhys07,HeSci14},
and nanoscale domains of similar PG magnitudes are spontaneously formed. The average PG magnitudes
are $\bar{\Delta}=31.4$ meV in optimally doped La-Bi2201 and $\bar{\Delta}=25.0$ meV in overdoped Pb-Bi2201,
showing the PG magnitude is decreased with the increased doping.
Since the O$_{\rm i}$(Sr) defects are dominant in our two Bi2201 samples,
we can collect unambiguous information by excluding the effects of the other two forms of O dopants.
In Figs.~\ref{fig:fign5}(a) and~\ref{fig:fign5}(b), the O$_\text{i}$(Sr) defects marked by
white dots are superimposed on top of the two PG maps. The O$_\text{i}$(Sr) defects
are more likely to be found in the red regimes (small PG magnitudes) than in the blue regimes (large
PG magnitudes), which indicates a negative correlation.

\begin{figure}[tp]
\centering
\includegraphics[width=0.55\columnwidth]{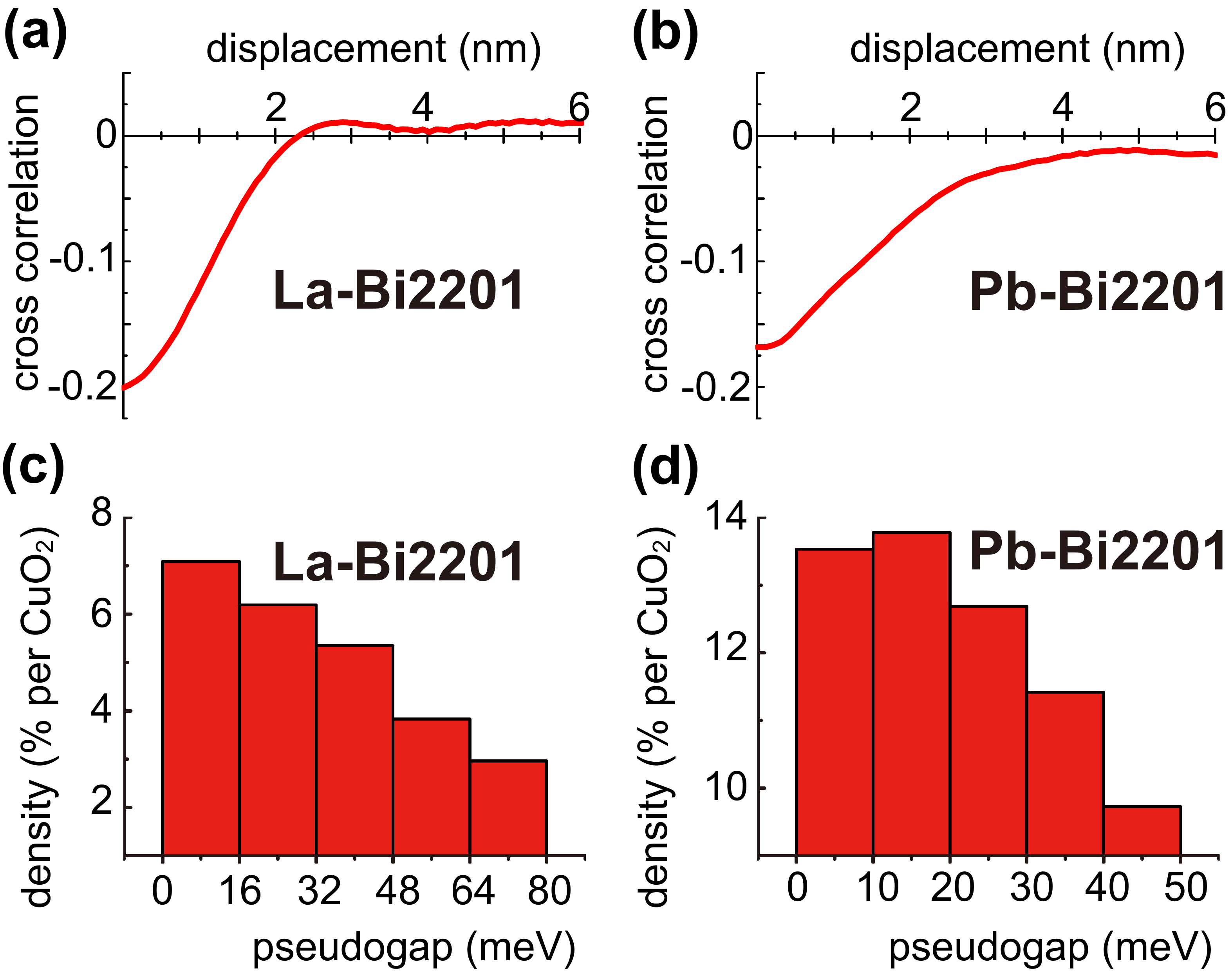}
\caption{(a-b) The cross correlation functions between the PG and distance maps %versus the spatial displacement $R$
for (a) La-Bi2201 and (b) Pb-Bi2201. (c-d) The histograms of the density of O$_\text{i}$(Sr) versus the PG magnitude for (c) La-Bi2201 and (d) Pb-Bi2201.
}
\label{fig:fign6}
\end{figure}

For a quantitative measurement, we introduce a
distance function of $D(\boldsymbol r)$, which is defined as the distance between the spatial position $\boldsymbol r$
and that of the nearest O$_\text{i}$(Sr) defect~\cite{ZeljkovicSci12}. A two-dimensional distance map of $D(\boldsymbol r)$
is obtained accordingly (see Figs.~\ref{fig:fign5}(c) and~\ref{fig:fign5}(d)).
For a given FOV, a normalized cross correlation between the PG and distance maps is given by
\be
C(\boldsymbol{R}) = -\frac{\int d^2\boldsymbol{r} [D(\boldsymbol{r})-\bar{D}][\Delta(\boldsymbol{r}+\boldsymbol{R})-\bar{\Delta}]}
{\sqrt{\int d^2\boldsymbol{r}[D(\boldsymbol{r})-\bar{D}]^2  \int d^2\boldsymbol{r}[\Delta(\boldsymbol{r})-\bar{\Delta}]^2  }},
\label{eq03}
\ee
where $\bar{D}$ is the average distance away from the nearest O$_\text{i}$(Sr) defect~\cite{ZeljkovicSci12}.
Equation~(\ref{eq03}) quantifies a dependence of the distance map reversely
on the PG map  with a spatial displacement $\boldsymbol R$. To further exclude the spatial randomness,
an integration over the angular coordinate is performed to give the cross correlation  $C(R)$
as a function of the spatial distance $R=|\boldsymbol R|$. The results of $C(R)$ for La-Bi2201 (FOV in Fig.~\ref{fig:fign1}(a))
and Pb-Bi2201 (FOV in Fig.~\ref{fig:fign1}(b)) are plotted in Figs.~\ref{fig:fign6}(a) and~\ref{fig:fign6}(b),
respectively. In both cases, the zero distance covariances $C(R=0)$ are around the value of $-0.2$
and the correlation lengths are in the order of nanometer. A negative cross correlation
prevails within a spatial range of interest, which is a characteristic feature of the O$_{\rm i}$(Sr) defects.

The zero distance covariance describes the dependence of the defect distance reversely on the PG magnitude. Another way
to interpret this number is to connect $C(R=0)$ with the average slope of $\rho(\Delta)$, which is the density of O$_{\rm i}$(Sr) defects
as a function of the PG magnitude. Accordingly, we divide the PG magnitudes of La-Bi2201 and Pb-2201 into
various ranges of $\Delta_j \le \Delta < \Delta_{j+1}$ and count the number of O$_{\rm i}$(Sr) defects within each $j$-th bin.
The defect density of $\rho_j=\rho(\Delta_j \le \Delta < \Delta_{j+1})$ is then calculated by dividing the defect number
over the area with the selected range of PG magnitudes. The histograms of $\rho_j$ are plotted in Figs.~\ref{fig:fign6}(c) and~\ref{fig:fign6}(d),
where a nearly monotonic decrease of $\rho_j$ with the PG magnitude is found for both samples.
The negative correlation between the O$_{\rm i}$(Sr) defects and the PG magnitudes is further confirmed.

In a previous STM study of Bi2212 samples, the influence of the O dopants to the PG state was explored~\cite{McElroySci05,ZeljkovicSci12}.
In the underdoped regime, the PG amplitude is positively correlated with the defect density for
all the three forms of the O dopants. In detail, the zero distance covariance $C(R=0)$ of the O vacancies
is almost twice that of the O$_{\rm i}$(Sr) defects, while this number becomes very small for
the O$_{\rm i}$(Bi) defects.  The PG state is thus considered to be mainly tuned by the O vacancies.
An interesting question raised is whether the other two types of O dopants really enhance the PG magnitude or not,
which could be hidden due to the strong interplay with the O vacancies. Relatively speaking,
The O$_{\rm i}$(Sr) defects are closer to the CuO$_2$ layers than the O$_{\rm i}$(Bi) defects, and
should exhibit a stronger influence to the PG state~\cite{ZhouPRL07}.
In the optimally doped Bi2212 sample
with a negligible density of the O vacancies, the previous experiment showed a negative correlation
between the O$_{\rm i}$(Sr) defects and the PG magnitude, although the O$_{\rm i}$(Bi) defects are still positively
correlated with the PG magnitude. With the O$_{\rm i}$(Sr) defects dominate in optimally
doped La-Bi2201 and overdoped Pb-Bi2201, we further confirms the negative dependence that
the PG magnitude of BSCCO decreases with the increase of the O dopants, while this behavior could be ambiguous
due to coexistence of different forms of O dopants in Bi2212.

\section{Summary}
\label{sec4}

In this paper, we apply the STM technique to study oxygen dopants in an optimally doped La-Bi2201 and
an overdoped Pb-Bi2201 samples from several different perspectives.
The characteristic features in the differential conductance spectrum enable us to distinguish three
different forms of O dopants: the interstitial O defects on the SrO layer (O$_{\rm i}$(Sr)),
the interstitial O defects on the BiO layer (O$_{\rm i}$(Bi)), and the O vacancies on the SrO layer.
The spatial distributions of these three forms of O dopants are determined by the conductance maps
at different characteristic voltages. In both La-Bi2201 and Pb-Bi2201 samples, the
number of O$_\text{i}$(Sr) defects is dominant, as compared to those of the other two forms.
A key parameter, the hole carrier density is estimated using the measured density of the O$_\text{i}$(Sr)
defects. This estimation is consistent with the Luttinger count extracted from the FS structure for
overdoped Pb-Bi2201. A large difference between the two estimations is however observed in optimally
doped La-Bi2201, which requires a further theoretical calculation. Subsequently, the
microscopic dependence of the PG state on the O$_\text{i}$(Sr) defects is explored.
Through a precise location of the O$_\text{i}$(Sr) defects on the PG maps, we determine a negative
correlation in both samples from an averaged cross correlation function and the histograms
of the O$_\text{i}$(Sr) densities. Our experimental measurement thus verifies an earlier observation of
the O$_\text{i}$(Sr) defects in optimally doped Bi2212, which is however complicated by the
dominant positive correlation of the O vacancies in underdoped Bi2212. The systematic researches
of both Bi2212 and Bi2201 samples can eventually reveal the microscopic mechanism of electronic
orders in the family of BSCCO superconductors.

\begin{acknowledgments}
This work is supported by the National Basic Research Program
of China (2015CB921004, 2014CB921203), the National Natural Science Foundation of
China (NSFC-11374260), and the Fundamental Research Funds
for the Central Universities in China (2016XZZX002-01). XJZ thanks financial support
from the National Natural Science Foundation of
China (NSFC-11334010), the National Key Research and Development Program of
China (2016YFA0300300), and the Strategic Priority Research Program (B) of the
Chinese Academy of Sciences (XDB07020300).
\end{acknowledgments}


\begin{thebibliography}{56}
\bibitem{ref}
We make the same caculations with the Bi2212 data extracted from Refs. [7] and [24] and obtain the same behavior.
\end{thebibliography}
\end{document}